\documentclass[doublecol]{epl2}

\usepackage{graphicx}
\usepackage{amsmath}
\usepackage{amsfonts}
\usepackage{amssymb}
\usepackage{enumerate}
\usepackage{longtable}
\usepackage{dcolumn}
\usepackage{bbm}
\usepackage{bm}

\newcommand {\e}  {\mathrm{e}}
\newcommand {\im} {\mathrm{i}}
\newcommand{\ipa}{IPA-CuCl$_3$ }

\newcommand{\ket}[1]{\left| #1 \right>}

\DeclareMathOperator{\dx}{d\!}

\begin{document}

\title{Microscopic Model for Bose-Einstein Condensation and Quasiparticle Decay}

\author{Tim Fischer\thanks{fischer@fkt.physik.tu-dortmund.de},
Sebastian Duffe and
G\"otz S. Uhrig\thanks{goetz.uhrig@tu-dortmund.de}}

\institute{Theoretische Physik I, Technische Universit\"at Dortmund, 
Otto-Hahn Stra\ss{}e 4, 44221 Dortmund, Germany}


\abstract{Sufficiently dimerized quantum antiferromagnets display elementary 
  $S=1$ excitations,
  triplon quasiparticles, protected by a gap at low energies.
  At higher energies, the triplons may decay into two or more
  triplons. A strong enough magnetic field induces Bose-Einstein condensation of
  triplons. For both phenomena the compound \ipa is
  an excellent model system. Nevertheless no quantitative model was determined 
  so far despite numerous studies. Recent theoretical progress 
  allows us to analyse
  data of inelastic neutron scattering (INS) and of magnetic susceptibility
  to determine the four magnetic couplings $J_1\approx -2.3 \text{ meV}$, 
  $J_2\approx 1.2\text{ meV}$, $J_3\approx 2.9\text{ meV}$ and 
  $J_4\approx -0.3\text{ meV}$.
  These couplings determine IPA-CuCl$_3$ as system of coupled asymmetric 
  $S=1/2$    Heisenberg ladders
  quantitatively. The magnetic field dependence of the lowest modes in the 
  condensed phase
  as well as the temperature dependence of the gap without magnetic field 
  corroborate this microscopic model.
}


\maketitle

Low-dimensional antiferromagnetic quantum spin systems display various
fascinating properties, e.g., spin-Peierls transition \cite{bray75,hase93a},
appearance of a Haldane gap for integer spins
\cite{halda83,renar87}, 
high-temperature superconductivity upon doping \cite{bedno86},
and the Bose-Einstein condensation (BEC) in spin-dimer systems 
\cite{affle91,shira97,giama99,garle07}, where the latter one is characterized
by a phase transition from a non-magnetic phase to a long-range
antiferromagnetically ordered gapless phase at a critical magnetic field
$H_{c1}$.

Another fascinating phenomenon recently observed in 
low-dimensional antiferromagnets is the decay of their
elementary $S=1$ excitations, triplons \cite{schmi03c}, 
at higher energies so that the triplons exist only in 
a restricted part of the Brillouin zone \cite{stone06,masud06}.
Theoretically as well,
there is rising interest in the understanding and quantitative
description of this phenomenon for gapped triplons 
\cite{kolez06,zhito06a,bibik07,fisch10a}
as well as for gapless magnons \cite{zhito99,zheng06a,chern09}.

The description of quasiparticle decay faces an intrinsic difficulty.
The merging of the long-lived, infinitely sharp elementary triplon
with a multitriplon continuum requires to describe the resulting resonance 
and its edges precisely. This is still a challenge for numerical approaches
such as exact diagonalization or dynamic density-matrix renormalization
\cite{kuhne99a}.
Diagrammatic approaches are able to capture the qualitative features 
but may encounter difficulties in the quantitative description 
in the regime of strong merging where the sharp mode dissolves
completely in the continuum because this is a strong coupling
phenomenon \cite{kolez06,zhito06a}. 
Unitary transformations also face difficulties
when modes of finite life-time occur \cite{fisch10a}.

A crucial step in the understanding of both phenomena is
to identify a suitable experimental system.
The best studied candidate for the BEC in coupled spin-dimer systems is
TlCuCl$_3$. Unfortunately, recent research suggests that the high field
spectrum remains gapped \cite{sirke05,johan05}  in contrast to
what is expected from a phase where a continuous symmetry is broken.
This suggests the existence of anisotropies.
A promising alternative for a BEC in a spin-dimer system is 
(CH$_3$)$_2$CHNH$_3$CuCl$_3$ (isopropylammonium trichlorocuprate(II),
short: IPA-CuCl$_3$) where inelastic neutron scattering (INS)
 provides evidence for an almost exact realization of a BEC 
\cite{garle07,zhelu07}.

A suitable experimental system to study
triplon decay in detail is searched for. The two-dimensional (2D) PHCC 
\cite{stone01,stone06} is a candidate, but it involves 
eight different couplings so that a quantitative characterization is
impossible to date.
Due to its quasi one-dimensional (quasi 1D) structure, IPA-CuCl$_3$ is again
a more promising candidate. This compound seems to realize the theoretically
proposed situation for BEC in coupled spin ladders  \cite{giama99}.

But in spite of many years of intensive studies
\cite{rober81,manak97,manak00b,masud06,manak07b,garle07,hong10a}
no quantitative microscopic model for IPA-CuCl$_3$ is
established. The present work aims at filling this gap.
Theoretically, our study is based on continuous unitary transformations 
(CUTs) of models with quasiparticle decay \cite{fisch10a}
and on high temperature series expansions for asymmetric spin ladders
which are topologically equivalent to dimerized and
frustrated spin chains \cite{buhle01a}. The experimental input used
in INS data \cite{masud06} and magnetic susceptibility $\chi(T)$ data
\cite{manak97}. We will illustrate why it is intrinsically difficult
to determine the microscopic model.

Finally, we will compute the temperature
and the magnetic field dependence of the lowest magnetic modes as well
as the upper critical magnetic field $H_{c2}$, which induces full polarization.  
They all agree very well with experimental data 
\cite{garle07,zhelu07,manak08,zhelu08,nafra11} which supports
the advocated model.

\begin{figure}[htb]
    \begin{center}
     \includegraphics[width=0.87\columnwidth,clip]{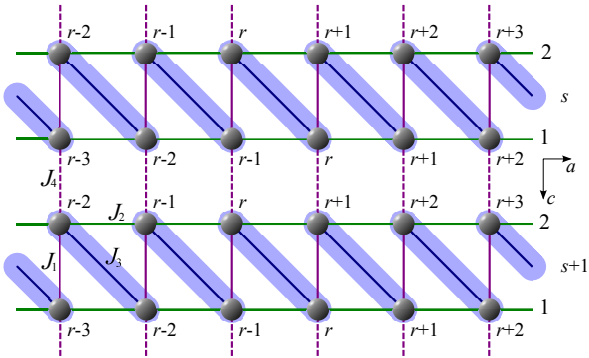}
    \end{center}
    \caption{\label{fig:ipa}(Color online) Sketch of
      IPA-CuCl$_3$. Circles indicate Cu ions with
      $S=1/2$. The couplings $J_1$ and $J_4$ are
      ferromagnetic ($J_1,J_4<0$) while $J_2$ and $J_3$ are
      antiferromagnetic ($J_2,J_3>0$). Two spins linked by 
      $J_3$ form a dimer.} 
\end{figure}

Since the   characterization of \ipa
 by Roberts \textit{et al.} \cite{rober81}
various spin models were discussed.
Manaka \textit{et al.} pointed out that 
the magnetic susceptibility of IPA-CuCl$_3$ can be explained by a
ferro-antiferromagnetically alternating Heisenberg $S=1/2$ chain with
ferromagnetic coupling twice as large as the antiferromagnetic
coupling \cite{manak97}. 
According to Hida \cite{hida92} the magnetic ground state is 
thus given by a gapped Haldane state \cite{halda83}.

The dispersions measured by INS \cite{masud06} and the crystal structure of
IPA-CuCl$_3$ indicates that the system is quasi-2D.
It is described by weakly coupled asymmetric spin $S=1/2$ Heisenberg 
ladders, see Fig.\ \ref{fig:ipa}, with
\begin{subequations}
\label{eq:ham_def}
  \begin{align}
    \label{eq:hamiltonian}
    H =& H_{\text{1D}} + H_{\perp} 
    \\
    \label{eq:hamiltonian_ladder}
    H_{\text{1D}} =& J_1 \sum_{r,s} \mathbf{S}_{1,r,s}\mathbf{S}_{2,r+1,s}
    + J_3 \sum_{r,s} \mathbf{S}_{1,r,s} \mathbf{S}_{2,r,s}  
    \notag \\
    +& J_2 \sum_{r,s} \left(\mathbf{S}_{1,r,s} \mathbf{S}_{1,r+1,s}+
    \mathbf{S}_{2,r,s} \mathbf{S}_{2,r+1,s} \right) 
    \\
    \label{eq:hamiltonian_int}
    H_{\perp} =& J_4\sum_{r,s} \mathbf{S}_{1,r,s} \mathbf{S}_{2,r+1,s+1}
  \end{align}
\end{subequations}
with two ferromagnetic couplings  $J_1$, $J_4<0$ and 
two antiferromagnetic couplings
 $J_2$, $J_3>0$. The dominant dimer coupling is $J_3$ so that we use the 
ratios $x=J_2/J_3$, $y=J_1/J_3$ and $z=J_4/J_3$. Let us first consider the 
ladders as isolated because the interladder coupling is small.
The standard view of these ladders takes the $J_3$ bonds to form the 
rungs of the ladder. Then $J_1$ is a diagonal bond. 

The key element of this model is the asymmetry 
of the spin ladders controlled by $J_1$. On the one hand,
the presence of $J_1$ spoils the reflection symmetry about
the center line of the ladder between the legs. This symmetry
would imply a conserved parity such that the triplons on the dimers
could be changed only by an even number \cite{knett01b,schmi05b}
so that no decay of a triplon into a pair of triplons could occur.
Hence the very presence of $J_1$ opens an important decay channel
for quasiparticle decay.

On the other hand, the two bonds
$J_2$ and $J_1$ represent the coupling of adjacent dimers.
Both contribute to the hopping of the triplons which is given 
in leading order by $2J_2-J_1$ \cite{uhrig96b}
while the interaction of 
adjacent triplons is proportional to $2J_2+J_1$. With
information only on the dispersion \cite{masud06}
it is impossible to determine $J_1$ and $J_2$ separately.
Hence, the same feature that induces the interesting quasiparticle
decay makes it particularly difficult to establish a microscopic model.

The BEC occurring in TlCuCl$_3$ was successfully described
by the bond-operator approach \cite{matsu02,matsu04}. But
this approach to spin-dimer systems is quantitatively reliable
only as long as the interdimer couplings $J_\mathrm{inter}$ are 
significantly smaller than the dimer coupling $J_\mathrm{dimer}$:
$|J_\mathrm{inter}| < J_\mathrm{dimer}/2$ \cite{norma11}. This limit
requires $|J_i| < J_3/2$ for $i\in\{1,2,4\}$ for \ipa which
does not hold \cite{manak97}. 
We will see below that $|J_1| \approx J_3$ provides
very good fits.

Thus we apply self-similar CUTs (sCUTs) to isolated ladders
\cite{mielk97b,reisc04,reisc06,kehre06}, 
modified to cope with decaying quasiparticles \cite{fisch10a}.
We use an infinitesimal generator which decouples the subspaces with zero or
one triplons from the remaining Hilbert space.
We can still decouple the 1-triplon subspace from 
the 2-triplon subspace for the isolated ladder.
The proliferating flow equations are truncated
if the range of the corresponding process exceeds
certain maximum extensions in real space
\footnote{The truncation scheme used for the Hamiltonian
is  $(d_2,d_3,\ldots, d_8)=(10,8,8,5,5,3,3)$ and for the observables
  $(d_1,d_2, \ldots, d_6)=(10,10,8,8,6,6)$,
where $d_j$ is the maximum extension for a process with $j$ creation 
and annihilation operators. Additionally, we keep only terms that 
create or annihilate at most $N=4$
triplons in the Hamiltonian and $N=3$ triplons in the observables,
see also Ref.\ \cite{reisc06,fisch10a}.}.
Thereby, the  ladders are mapped  to an effective model 
\begin{equation}
\label{eq:ham_1d_eff}
H_\text{1D,eff}=\sum_{h,l;\alpha}  \omega_0(h)
t_{\alpha,h,l}^{\dag}t_{\alpha,h,l}^{\phantom{\dag}}
\end{equation}
in terms of triplon creation $t_{\alpha,h,l}^{\dag}$ and annihilation operators
$t_{\alpha,h,l}^{\phantom{\dag}}$ in momentum space, where
$h$ is the wave vector component along the ladders, $l$ the one perpendicular
to  them, and $\alpha\in\left\{x,y,z\right\}$ the spin polarization.
These operators are the Fourier transforms of the bond operators 
\cite{chubu89a,sachd90} 
defined on the dimers in Fig.\ \ref{fig:ipa}.

The dispersion $\omega_0(h)$ depends only on $h$ because 
the CUT is applied to the isolated ladders which still
have to be coupled. This coupling is achieved in leading order
following the approach in Refs.\ \cite{uhrig04a,uhrig05a}.
The spin component $S^{\alpha}_{i,r,s}$ is taken as observable and
transformed into the new basis by the CUT. Then it reads
$S_{\text{eff},i,r,s}^{\alpha} := U^{\dag}  S_{i,r,s}^{\alpha} U$
\begin{equation}
    S_{\text{eff},i,r,s}^{\alpha}
    = \sum_{\delta} a_{i,\delta} ( t_{\alpha,r+\delta,s}^{\dag} +
    t_{\alpha,r+\delta,s}^{\phantom{\dag}} ) + \ldots \ ,
\end{equation}
where the dots stand for normal-ordered higher terms 
in the real space triplon operators
$t_{\alpha,r,s}^{\dag}$ ($t_{\alpha,r,s}$).
Knowing $S_{\text{eff},i,r,s}^{\alpha}$ allows us in a second step to 
write down the effective interladder coupling $H_\text{int,eff}$
in real space
\begin{eqnarray}
\nonumber
H_{\text{int,eff}} &=& J_4\sum_{r,s;\alpha}\sum_{\delta,\delta'}
    a_{1,\delta}a_{2,\delta'} [  t_{\alpha,r,s}^{\dag}
    (t_{\alpha,r+1+\left(\delta'-\delta\right),s+1}^{\dag} 
    \\
     &&  + t_{\alpha,r+1+\left(\delta'-\delta\right),s+1}^{\phantom{dag}})
+ \text{H.c.}].
\label{eq:ham_inter_eff}
\end{eqnarray}
This neglects trilinear and
higher contributions. The Fourier transform of $H_\text{int,eff}$
leads to 
$H_\text{eff}=H_\text{1D,eff}+H_\text{int,eff}$ amenable to
a Bogoliubov diagonalization yielding
\begin{subequations}
  \begin{align}
    \label{eq:hamilton-eff}
    H_\text{eff} =& \sum_{h,l;\alpha}  \omega(h,l) 
    b_{\alpha,h,l}^{\dag}b_{\alpha,h,l}^{\phantom{\dag}} 
    \\
    \label{eq:dispersion}
    \omega(h,l) =& \sqrt{\omega_0^2(h) + 4\omega_0(h) \lambda(h,l)} 
    \\
    \notag
    \lambda(h,l) =& - J_4
    \sum_{\delta,\delta'}a_{1,\delta}a_{1,\delta'}\cos\left(2\pi\left[ h
      \left(\delta + \delta' - 1 \right) - l \right]\right) 
  \end{align}
\end{subequations}
with bosonic operators $b_{\alpha,h,l}^{\dag}$ ($b_{\alpha,h,l}$).
In the Bogoliubov diagonalization the hardcore property of the 
bosons is neglected.
However this does not concern the large intraladder couplings, but
only the small interladder couplings so that the approach is still very
accurate \cite{exius10b}. The dispersion $\omega(h,l)$ 
makes a direct comparison with INS results possible.

\begin{table}[htb]
  \begin{center}
    \caption{\label{tab} 
      Parameters  for IPA-CuCl$_3$ compatible
      with INS \cite{masud06}}
    \begin{tabular}[c]{cccc}
      \hline \hline 
      $\ J_3$ [meV]\ & $\ x=J_2/J_3\ $ & $\ y=J_1/J_3\ $ & $\ z=J_4/J_3$\ \\
      \hline
      3.743 & 0.133 & -2.0 & -0.076 \\
      3.288 & 0.268 & -1.4 & -0.088 \\
      3.158 & 0.317 & -1.2 & -0.092 \\
      3.038 & 0.369 & -1.0 & -0.096 \\
      2.929 & 0.424 & -0.8 & -0.100 \\
      2.830 & 0.480 & -0.6 & -0.103 \\ 
      \hline
    \end{tabular}
  \end{center}
\end{table}

\begin{figure}[htb]
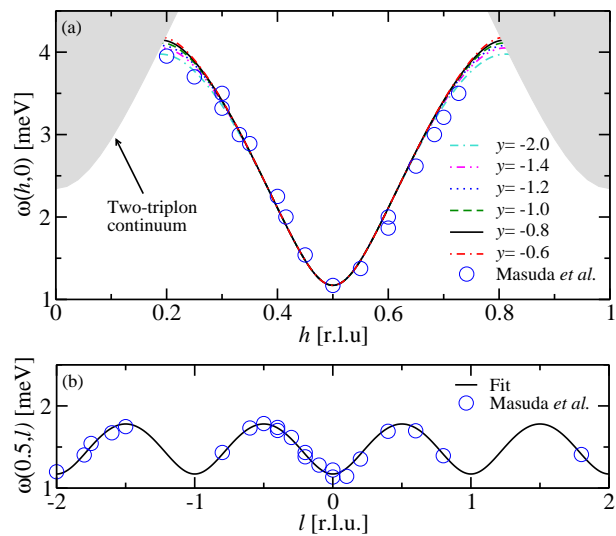

  \begin{center}
    \includegraphics[width=0.90\columnwidth,clip]{fig2a}
    \\[0.1cm]
    \includegraphics[width=0.90\columnwidth,clip]{fig2b}
    \\
  \end{center}
  \caption{\label{fig:disp} 
    (Color online) Circles are INS data  \cite{masud06}.
    (a) Dispersions $\omega(h,0)$ for $xyz$ triples in Tab.\ \ref{tab}. 
    The quasiparticle decay occurs where the dispersion 
    enters the 2-triplon continuum. 
    (b) Dispersion $\omega(0.5,l)$; all triples lead to coinciding curves.} 
\end{figure}

To determine the microscopic parameters we fix the value
$y=J_1/J_3$ and fit $x=J_2/J_3$, $z=J_4/J_3$,
and the energy scale $J_3$ to reproduce the experimental result 
(Eq.\ (2) in Ref.\ \cite{masud06})
\begin{eqnarray}
\nonumber
\omega(h,l)^2&=&a^2\cos^2(\pi h) + 
[\Delta^2+4b^2\sin^2(\pi l)]\sin^2(\pi h)
\\
&&+ c^2\sin^2(2\pi h)
\end{eqnarray}
with $a=4.08(9)$meV, $\Delta=1.17(1)$meV, $b=0.67(1)$meV and $c=2.15(9)$meV.
Thus, we obtain the triples $(x, y, z)$ in Tab.\ \ref{tab}. They all 
essentially imply  the same dispersion, see Fig.\  \ref{fig:disp}.
Hence, on the basis of the the INS data, 
one cannot decide which of the triples applies to IPA-CuCl$_3$.

The quasiparticle decay occurs where the dispersion 
enters the 2-triplon continuum. It does not prevent
to use the CUT for the isolated ladder since the
realistic parameters turn out to be such 
that the triplons do not decay \emph{without} the
interladder coupling. A quantitative description
of the decay is subject of ongoing research.

\begin{figure}[htb]
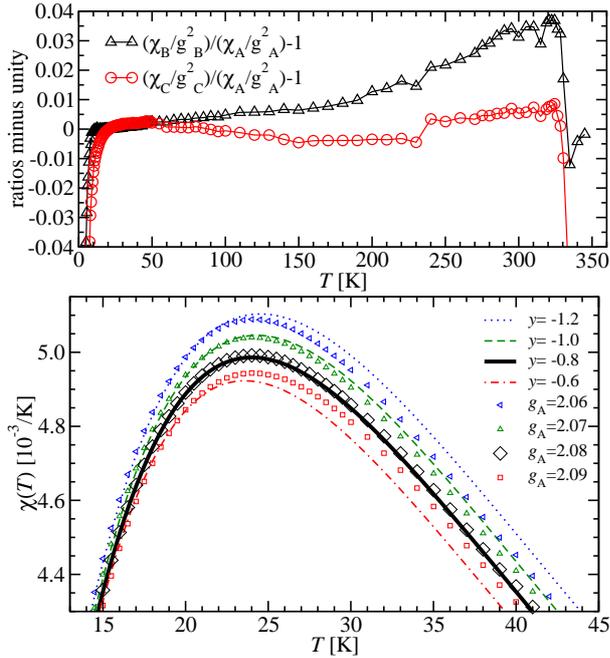

  \begin{center}
     \includegraphics[width=0.90\columnwidth,clip]{fig3a}
    \includegraphics[width=0.90\columnwidth,clip]{fig3b}
  \end{center}
  \caption{\label{fig:sus} 
    (Color online) Upper panel: Deviations of the
    experimental magnetic susceptibilities \cite{manak97} $\chi_\mathrm{B}$ and
    $\chi_\mathrm{C}$ in B and C direction relative
    to $\chi_\mathrm{A}$ for $g_\mathrm{A}=2.08$, $g_\mathrm{B}=2.06$,
    and $g_\mathrm{C}=2.25$, indicating anisotropies.
    Lower panel: Comparison of $\chi_\mathrm{A}(T)$ for various values 
    $g_{\text{A}}$ with theoretical results obtained by 
    Dlog-Pad\'e approximated high temperature series expansions
    for the $xyz$ triples from Tab.\ \ref{tab}.}
\end{figure}

In complement to the INS we use the temperature
dependence of the magnetic susceptibility $\chi(T)$ \cite{manak97}.
Starting from the spin isotropic Hamiltonian \eqref{eq:ham_def} 
the susceptibilities in different spatial direction have to be the same up to
scaling proportional to the squares of the Land\'e $g$-factors. This means
that $\chi_\mathrm{A} : \chi_\mathrm{B} : \chi_\mathrm{C}$ equals 
$g^2_\mathrm{A} : g^2_\mathrm{B} : g^2_\mathrm{C}$
where A, B, C indicate the directions normal to the corresponding
surfaces of the crystal \cite{manak97}.
Fig.\ \ref{fig:sus}a displays that the three susceptibilities
can be scaled to coincide for $g_\mathrm{A}=2.08$, $g_\mathrm{B}=2.06$,
and $g_\mathrm{C}=2.25$ within about 3\%.
This choice of $g$-factors fulfills the experimental constraints 
\cite{manak97,manak00b} $g_\mathrm{A}, g_\mathrm{B} \in [2.06,2.11]$ and 
$g_\mathrm{C}=2.25 -2.26$  best. We conclude that an spin isotropic 
Hamiltonian such as \eqref{eq:ham_def} provides
a very good description, although anisotropies, e.g., Dzyaloshinskii-Moriya
terms, can be present with a relative size of a few percent.
This agrees with findings from electron paramagnetic resonance \cite{manak00b}.

Theoretically, we use the high temperature series expansion for
the isolated asymmetric ladder \cite{buhle01a} providing 
 series in $\beta=1/T$ up to order $\beta^{n+1}$ with $n=10$ 
 denoted by $\chi_\text{1D}$. The 2D series $\chi_\text{2D}$
 obeys the relation $\chi_\text{2D}^{-1}=\chi_\text{1D}^{-1}+J_4$ in 
interladder mean-field  approximation, i.e., in leading order in 
$J_4$. We use standard Dlog-Pad\'e approximation \cite{domb89} 
to deduce the full $\chi(T)$ from $\chi_\text{2D}$ and from the asymptotic 
behavior $\chi_\text{2D}(\beta) \propto \beta^0 \exp\left(-\Delta \beta \right)$ 
for  $1/\beta \ll \Delta$.  The
result\footnote{All theory curves rely on the [7,4] Dlog-Pad\'e approximant
in $u=\beta/(1+\beta)$. Data from other Dlog-Pad\'e approximants, e.g., [9,2], 
agrees  within line width except at very low temperatures.} is plotted
in Fig.\ \ref{fig:sus} and compared to $\chi_\mathrm{m}$
measured in [emu/g] and converted according to
$\chi(T)  = {m_{\text{mol}} k_{\text{B}}}{\left(g \mu_{\text{B}}\right)^{-2}   
N_{\text{A}}^{-1}} \chi_{\text{m}}(T)$.
Here $m_{\text{mol}}$ is the molar mass of IPA-CuCl$_3$, $k_{\text{B}}$ the
Boltzmann constant, $\mu_{\text{B}}$ the Bohr magneton and $N_{\text{A}}$ the
Avogadro constant. 

Fig.\ \ref{fig:sus}b illustrates that
theory and experiment agree indeed best for $g_\mathrm{A}=2.08$
and the triple of $y=-0.8$. As an asset, we stress that even without
the value of $g_\mathrm{A}$, the \textit{position} and the 
\textit{shape} of the  maximum of $\chi(T)$ fits best for the triple of $y=-0.8$
and one can deduce \textit{deduce} that the $g_\mathrm{A}$-factor
is around $2.08$. As a caveat, we stress the very weak
dependence of $\chi(T)$ on $y$ in a triple tuned to the INS data. 
By assuming $g_{\text{A}}=2.08\pm 0.01$ we estimate the error
of our analysis to be
$x=0.42 \pm 0.06$, $y=-0.8 \pm 0.2$ and $z=-0.100 \pm 0.004$
implying $J_1= -2.3 \pm 0.6 \text{ meV}$,
$J_2= 1.2 \pm 0.2\text{ meV}$,
$J_3= 2.9 \pm 0.1 \text{ meV}$ and $J_4= -0.292 \pm 0.001\text{ meV}$.
These values establish the microscopic model for
IPA-CuCl$_3$. We highlight that the ferromagnetic coupling $J_1$
does not dominate over the antiferromagnetic coupling $J_3$
because $|y|\lessapprox 1$, 
in contrast to the previous purely 1D analysis \cite{manak97}.

The derived microscopic model successfully 
passes three checks: The BEC is well-described, the upper critical 
field $H_{c2}$ 
agrees to experiment and the temperature dependence 
of the spin gap matches recent data.

First,we follow Refs.\ \cite{somme01,matsu02,matsu04} to describe
the BEC and perform the local transformation
\begin{subequations}
  \label{eq:magnetization_trafo}
  \begin{align}
    \ket{\tilde{s}_{\mathbf{r}}}&= u \ket{s_{\mathbf{r}}} + v \e^{\im
     \mathbf{Q}_0 \mathbf{r}} \left(f \ket{{t}_{+,\mathbf{r}}}+g
      \ket{{t}_{-,\mathbf{r}}}\right) \\
    \ket{\tilde{t}_{+,\mathbf{r}}}&= u \left(f \ket{{t}_{+,\mathbf{r}}}+g
      \ket{{t}_{-,\mathbf{r}}}\right) - v \e^{\im
     \mathbf{Q}_0 \mathbf{r}} \ket{s_{\mathbf{r}}} \\
    \ket{\tilde{t}_{0,\mathbf{r}}}&= \ket{{t}_{0,\mathbf{r}}} \\
    \ket{\tilde{t}_{-,\mathbf{r}}}&= f \ket{{t}_{-,\mathbf{r}}}-g
      \ket{{t}_{+,\mathbf{r}}}
  \end{align}
\end{subequations}
in real space with $u=\cos(\theta)$, $v=\sin(\theta)$, $f=\cos(\varphi)$
and $g=\sin(\varphi)$, the position $\mathbf{r}=(r,s)$ and the wave vector 
$\mathbf{Q}_0=(\pi,0)$ of the minimum of the dispersion. 
The triplon states $\ket{{t}_{m}}$ with $m\in\left\{-,0,+\right\}$ are given by
 $\ket{{t}_{-}}=1/\sqrt{2}\left(\ket{{t}_{x}}-\im
  \ket{{t}_{y}}\right)$, $\ket{{t}_{0}}=\ket{{t}_{z}}$ and 
$\ket{{t}_{+}}=1/\sqrt{2}\left(\ket{{t}_{x}}+\im
 \ket{{t}_{y}}\right)$. The tensor
product of all singlet states $\ket{s_{\mathbf{r}}}$ is the vacuum $\ket{0}$,
so that the hardcore triplon creation operator with $m\in\{-,0,+\}$ is
defined by  $\tilde t_{m,\mathbf{r}}^\dag \ket{0} := 
\ket{\tilde{t}_{m,\mathbf{r}}}$ and the annihilation by
$\tilde t_{m,\mathbf{r}} \ket{\tilde{t}_{m,\mathbf{r}}} :=\ket{0}$
and so on. In this basis the magnetic field is described by the operator
$-h(t_+^\dag t_+^{\phantom{\dag}} - t_-^\dag t_-^{\phantom{\dag}} )$. The two
independent variables $\theta$ and $\varphi$ are varied to minimize the
classical ground state energy. This choice also ensures  that 
(i) all linear terms in the triplon operators vanish and 
(ii) a massless Goldstone mode appears as it has to be.

\begin{figure}[htb]
    \begin{center}
     \includegraphics[width=0.90\columnwidth,clip]{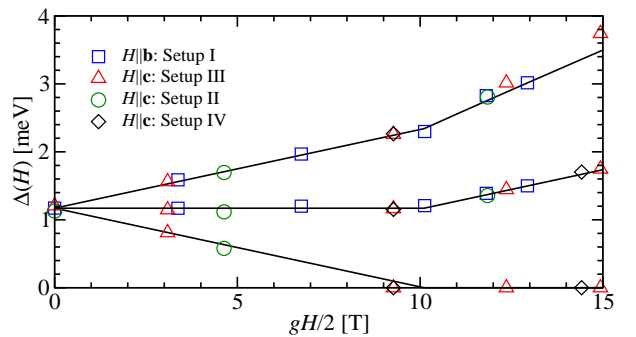}
    \end{center}
    \caption{\label{fig:gap_H} 
    	(Color online) Gaps in IPA-CuCl$_3$ vs.\
      the reduced magnetic field $gH/2$. Solid lines show theoretical results, 
      see main text; symbols mark experimental data from
      Refs.\ \cite{zhelu07} (setup I \& IV) and \cite{garle07} 
      (setup II \& III).} 
\end{figure}

Previous work \cite{somme01,matsu02,matsu04} applied the transformation
\eqref{eq:magnetization_trafo} to the original spin model.
This is not possible for \ipa because the dimers are too strongly coupled.
Hence the CUT is mandatory and we apply the real
space transformation \eqref{eq:magnetization_trafo} to 
$H_\mathrm{1D,eff}+H_\mathrm{inter,eff}$ from 
Eqs.\ (\ref{eq:ham_1d_eff},\ref{eq:ham_inter_eff})
keeping the bilinear terms. 
Fourier transformation and Bogoliubov diagonalization
finally provides the lowest lying modes. Their resulting gap energies
are displayed in Fig.\ \ref{fig:gap_H}.
\emph{No}  parameters are adjusted.

Second, the upper critical field $H_{c2}$ 
can be determined exactly for the spin model \eqref{eq:hamiltonian} to be
$H_{c2}=(2J_2+J_3)/(g\mu_{\text{B}}) \approx 2/g \cdot 45.8 \text{
  T}$. After the transformation \eqref{eq:magnetization_trafo} is applied
  to the dispersion obtained from CUT we obtain 
$H_{c2} \approx 2/g \cdot 45.1 \text{ T}$. 
The very good agreement of these two values strongly
supports the approximations made.
Additionally, the theoretical values also match 
the experimental result \cite{manak08} $H_{c2}=(43.9\pm 0.1) \text{
T} (2/g)$ within 4\%. In view of the neglect of anisotropies and
magnetoelastic effects, cf.\ Ref.\ \cite{johan05}, 
this nice agreement lends independent support to the 
advocated microscopic model.

\begin{figure}[htb]
    \begin{center}
     \includegraphics[width=0.90\columnwidth,clip]{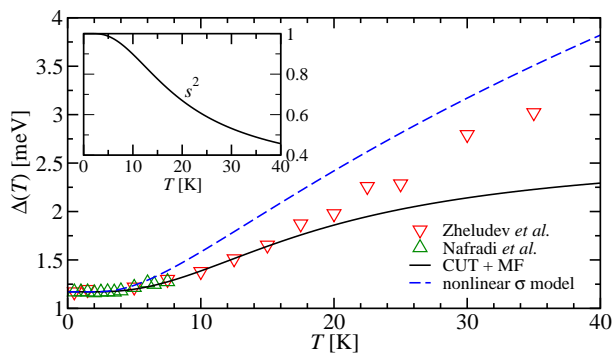}
    \end{center}
    \caption{\label{fig:gap_T} 
    (Color online) Spin gap in IPA-CuCl$_3$ vs.\
      temperature $T$. Lines show theoretical results from CUTs and 
      mean-field (solid) and from the nonlinear $\sigma$ model (dashed), 
      respectively. Inset: Temperature dependence of the condensate 
      fraction $s^2$.} 
\end{figure}

Third, the temperature dependence of the gap
$\Delta(T)$ supports that the low-lying excitations are
hardcore triplons. We apply the mean-field approach in 
Refs.\ \cite{sachd90,troye94,ruegg95,exius10a,norma11}
to $H_\mathrm{1D,eff}+H_\mathrm{inter,eff}$
from Eqs.\ (\ref{eq:ham_1d_eff},\ref{eq:ham_inter_eff}).
In each nonlocal term ($t^\dag_{m,\mathbf{r}}t^{\phantom{\dag}}_{m,\mathbf{r}'}$ or
$t^\dag_{m,\mathbf{r}}t^{{\dag}}_{-{m},\mathbf{r}'}$ or
$t^{\phantom{\dag}}_{m,\mathbf{r}}t^{\phantom{\dag}}_{-{m},\mathbf{r}'}$
with $\mathbf{r}
\neq \mathbf{r}'$) all creation operators $t^\dag_{m,\mathbf{r}}$ are
multiplied by the singlet annihilation $s_{\mathbf{r}}$ and
the annihilation operators $t_{m,\mathbf{r}}$  by the singlet creation 
$s^\dag_{\mathbf{r}}$. Local terms remain unchanged
because they do not change the local singlet number.
Finally all singlet operators are
replaced by the condensate value $s(T) = \left\langle
s^\dag \right\rangle = \left\langle  s\right\rangle$  with 
$s\in \left[0,1\right]$.
In a nutshell, a factor $s^2$ appears in front of each nonlocal term.

This implies a dependence of the dispersion on $s$ and hence on 
temperature \cite{troye94,exius10a,norma11}, denoted by $\omega_{s(T)}(h,l)$. 
The self-consistent solution is found from
the hardcore condition $1=\langle s^\dag_{\mathbf{r}} s_{\mathbf{r}}
+\sum_m t^\dag_{m,\mathbf{r}}t^{\phantom{\dag}}_{m,\mathbf{r}}\rangle$
leading to $s^2(T) = 1 -  3z/(1+3z)$ with
\begin{equation}
  z = \int_{-1/2}^{1/2} \dx h
  \int_{-1/2}^{1/2} \dx l\ \e^{-\beta\omega_{s(T)}(h,l)} .
\end{equation}
Figure \ref{fig:gap_T}
compares the result (solid line) of this simple approximation
to INS data \cite{zhelu08,nafra11}. 
Up to $15$ K the experimental data is matched perfectly. 
We attribute the discrepancy at higher temperatures
to the insufficient treatment of the hardcore constraint by
the above approach (for $15$ K the condensate fraction $s^2$ is only $0.77$). 
Note that we only apply the mean-field theory to the 
dispersion obtained from CUT, not to the original spin model
as done previously \cite{ruegg95,norma11} because
\ipa is not far enough in the dimer limit.

For comparison, we also include
$\Delta(T)$ as derived from the nonlinear $\sigma$ model on
1-loop level \cite{senec93} in Fig.\ \ref{fig:gap_T} (dashed line).
It is obtained from
\begin{equation}
  \label{eq:nlsm}
  C=\int_{-1/2}^{1/2} \dx h \int_{-1/2}^{1/2} \dx l\ 
  \frac{\coth\left(\beta\omega(h,l,T)/2 \right)}{\omega(h,l,T)} 
\end{equation}
with $\omega(h,l,T):=\sqrt{\omega^2(h,l) + \Delta^2(T) - \Delta^2(0)}$; 
the constant $C$ is determined for $T=0$. 
Interestingly, this approach describes the
experimental data less accurately if the experimental dispersion at $T=0$
is used for $\omega(h,l)$, cf.\ Ref.\ \cite{nafra11}. 
We presume that the hardcore constraint is not accounted for sufficiently 
well by Eq.\ \eqref{eq:nlsm}.

In summary, we showed that the available experimental evidence
for \ipa
is consistent with a quantitative model of weakly coupled
asymmetric $S=1/2$ spin ladders with hardcore triplons as excitations. 
Such systems are of great current interest because they 
allow for the study of Bose-Einstein condensation of triplons 
and of the massless
excitations above this condensate \cite{giama99,garle07}.
Additionally, they  represent gapped quantum liquids known to display 
considerable quasiparticle decay \cite{kolez06,zhito06a,bibik07,fisch10a}.

Our high-precision analyses of  inelastic neutron scattering data
 and of the temperature dependence of the magnetic susceptibility
is based on advances in continuous unitary transformations
\cite{fisch10a} and high temperature series expansions. 
The established quantitative model paves the way
for further quantitative studies, both experimental and theoretical,
of the decay of massive quasiparticles and of the 
condensation of hardcore bosons. 

The latter is illustrated by the excellent agreement of
the calculated gap energies as function of magnetic field.
Additionally, the description of IPA-CuCl$_3$ by dispersive hardcore triplons
is strongly supported by the agreement of the
temperature dependence of the spin gap.

By this work, a quantitative model for IPA-CuCl$_3$ is
established. Concomitantly, we exemplarily showed 
how CUT results in one dimension at zero
temperature and zero magnetic field can be extended to
render a quantitative description in two dimensions at
finite temperature and finite magnetic field possible. 
We expect this approach to continue to be fruitful also
for other systems.

\acknowledgments

We thank T.~Lorenz, O.P.~Sushkov, H.~Manaka and A.~Zheludev for insightful
discussions and the latter two and B.~N\'afr\'adi  
for providing experimental data.  
This work was supported by the NRW Forschungsschule
``Forschung mit Synchrotronstrahlung in den Nano- und Biowissenschaften''
and by the DAAD.


\end{document}